\documentclass[12pt]{spieman}  

\usepackage{amsmath,amsfonts,amssymb}
\usepackage{comment}
\usepackage{graphicx}
\graphicspath{ {./Figures/} }
\usepackage{setspace}
\usepackage{tocloft}
\usepackage[font={large}]{caption}

\usepackage{ctable} 
\newcommand{\vk}{von-K\'{a}rm\'{a}n}
\title{A generalized approach to compensate for low and high frequency errors in FFT based phase screen simulations}

\author[a]{Sorabh Chhabra}
\author[a,e]{Jyotirmay Paul}
\author[a, c, d]{A. N. Ramaprakash}
\author[b]{Avinash Surendran}
\affil[a]{Inter-University Center for Astronomy And Astrophysics, Pune 411007, India}
\affil[b]{W. M. Keck Observatory,Kamuela, Hawaii 96743, USA}
\affil[c]{Cahill Center for Astronomy and Astrophysics, California Institute of Technology, Pasadena, CA 91125, USA}
\affil[d]{Institute of Astronomy, Foundation for Research and Technology-Hellas, Voutes, 71110 Heraklion, Greece}
\affil[e]{University of Liège, Space Sciences Technologies \& Astrophysics Research Institute, Liège 4000, Belgium}

\cftpagenumbersoff{figure}
\cftpagenumbersoff{table}

\begin{document} 
\maketitle

\begin{abstract}
Fast Fourier Transform based phase screen simulations give accurate results only when the screen size ($G$) is much larger than the outer scale parameter ($L_0$). Otherwise, they fall short in correctly predicting both the low and high frequency behaviours of turbulence induced phase distortions. Sub-harmonic compensation is a commonly used technique that aids in low-frequency correction but does not solve the problem for all values of screen size to outer scale parameter ratios $(G/L_0$). A subharmonics based approach will lead to unequal sampling or weights calculation for subharmonics addition at the low-frequency range and patch normalization factor. We have modified the subharmonics based approach by introducing a Gaussian phase autocorrelation matrix that compensates for these shortfalls. We show that the maximum relative error in structure function with respect to theoretical value is as small as 0.5-3\% for $(G/L_0$) ratio of 1/1000 even for screen sizes up to 100 m diameter.
\end{abstract}

\keywords{Phase Screen, Fast Fourier Transform, Subharmonic, Autocorrelation, Phase structure function}

{\noindent \footnotesize\textbf{*}Sorabh Chhabra,  \linkable{sorabh@iucaa.in} / \linkable{sorabh.chhabra@gmail.com} }

\begin{spacing}{2}   

\section{Introduction}
\label{sec:intro}  
\begin{figure}[h!]
    \centerline{\includegraphics[width=0.75\textwidth]{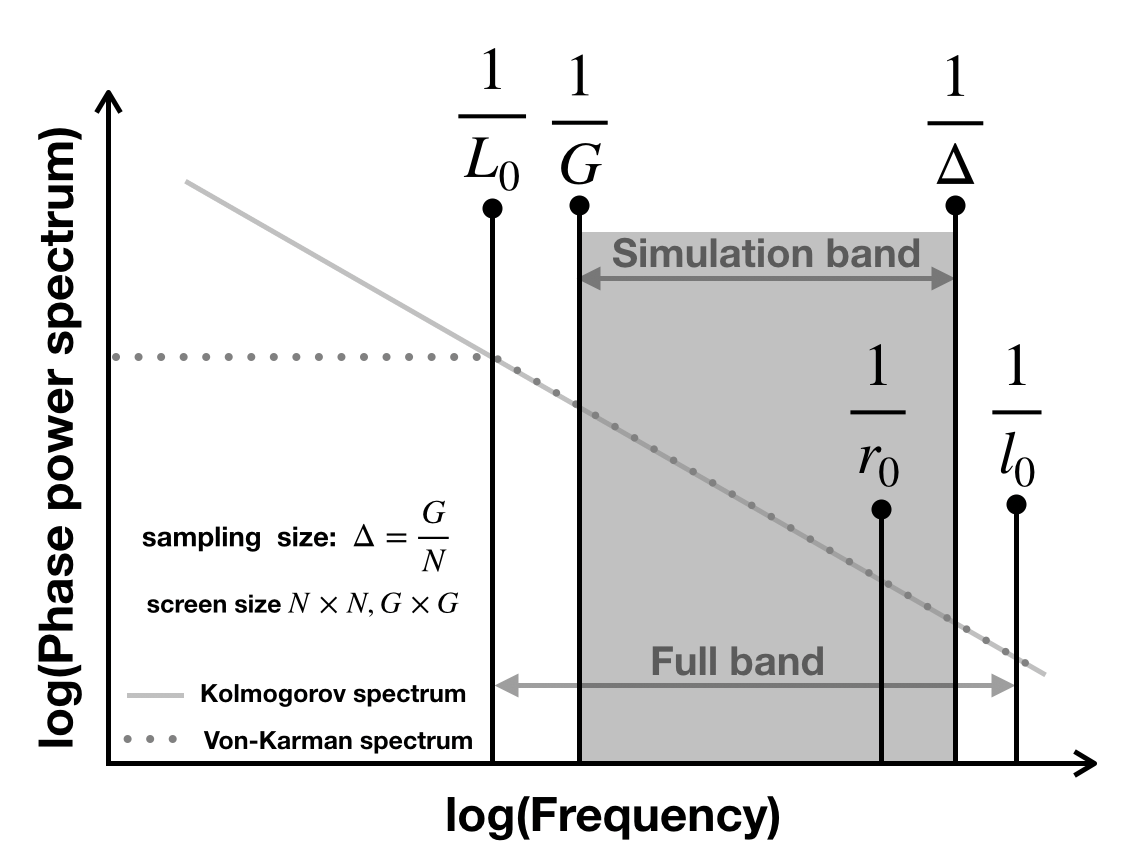}}
    \caption{\label{fig:band} Comparison between Simulation band and Full band}
\end{figure}

 Accurately simulating the atmospheric turbulence
behaviour is well recognized as very challenging. For a variety of purposes such as the design and development of adaptive optics systems, speckle imaging techniques, atmospheric propagation studies etc., it is essential to simulate good atmospheric phase screen models.  Methods based on Zernike polynomial expansions\cite{Roddier90}, FFT-based methods \cite{Herman90,Johansson94,Sedmak98,McGlamery76,Sedmak04,Xiang14,Xiang12}, Low Frequency Optimization method \cite{Zhang19} etc. have been in use for this purpose. The Zernike polynomial method, which is widely in use, has a limitation due to the maximum number of coefficients needed for accurate compensation. The optimization method which compensates accurately for low frequency part of the spectrum by using unequal sampling and unequal weight in low frequency region, does not cover high frequency deficiencies.  Among these, FFT-based methods are computer memory size friendly and widely accepted. But, FFT operators assume uniform sampling for the non-uniformly distributed phase power spectrum which can lead to underestimation in the low and high frequency out of band regions, as illustrated in Fig.~\ref{fig:band}. Thus, it has limitations in recreating the true phase power spectrum. To compensate for low-frequency components, Johansson and Gavel \cite{Johansson94} suggested employing the modified subharmonics equation (an adaptation from Lane et al. \cite{Lane92}), which works well up to an infinite outer scale length. Sedmak \cite{Sedmak04} later compared the performance of this method with that of Lane et al. \cite{Lane92} by actually calculating the phase structure function from the simulated screen. He improved upon Lane et al. \cite{Lane92} by employing different fine tuned subharmonic weights for different $G/L_0$ ratios. Results from his analysis show that these FFT-based simulations are accurate for large screen size ($G$) to outer scale parameter ($L_0$) ratios. For a screen size of $G$ = 200 m and outer scale of $L_0$ = 25 m, the maximum relative error in the simulation approaches 1\%. Our simulations demonstrate that the errors from low-frequency components start shooting up once we move to smaller $G/L_0$ ratios, even after compensating with modified subharmonics.

In Fig.~\ref{fig:band} we illustrate \cite{sedmak_private} a typical situation where the simulation band $\big(\frac{1}{G}-\frac{1}{\Delta}\big)$ is actually  smaller than full band $\big(\frac{1}{L_0}-\frac{1}{l_0}\big)$, where $\Delta$ is the sampling size defined as the ratio of screen size $G$ to sampling number $N$ and $l_0$ is the inner scale parameter. In practice, the simulations are often curtailed at the low frequency end, to a few times the optical beam size (say as determined by the telescope or laser beam diameter), while at the high frequency end, they often extend to only a few times that determined by the Fried parameter $r_0$.  Clearly, the larger the simulation band to full band ratio, the more accurate the simulated results will be. 

On the one hand, the apertures of upcoming and future astronomical telescopes are often of the same order or even larger than the typical median outer scale sizes of about 20 m - 25 m\cite{ziad2016review}. On the other hand, wavefront sensing and compensation technologies are fast progressing that Nyquist sampling at $r_0$ scales even for large aperture telescopes are becoming quite possible. Thus atmospheric turbulence simulations have to deal with a wide range in a multi-dimensional parameter space.
 
For working with very small apertures relative to the outer scale, it may appear that we need to simulate only a relatively small screen size. But cutting out small apertures from a larger screen introduces deviation from phase structure function due to misrepresentation of low frequency components present in the small screen power spectrum.

In this paper, we present an approach and a corresponding algorithm to deal with  phase screen simulations for a wide range of $G/L_0$ ratios, using the FFT-based method. Our technique builds upon the modified subharmonic approach of Johansson and Gavel \cite{Johansson94} and is inspired by Jingsong Xiang's \cite{chhabra2020gaussian}. It works well for space- and time-invariant, zero intermittency atmospheric turbulence. Section~\ref{sec:phase_production} explains how to obtain phase autocorrealtion matrix using phase power spectrum, 
Section~\ref{sec:algorithm} presents the algorithm part to compensate for the remaining error in phase structure function calculation,
Section~\ref{sec:flow_chart} steps through the implementation of the algorithm with the help of a flow chart,
Section~\ref{sec:Validation} covers the validation of the technique using results from simulated phase screens,
and Section~\ref{sec:Conclusion} provides the concluding remarks.

\section{Obtaining phase autocorrelation matrix using phase power spectrum}
\label{sec:phase_production}

The 2D phase structure function and phase autocorrelation matrix are related as follows\cite{roddier1999adaptive} :
\begin{equation}
\label{eq:eq1}
    D_{\phi}(m,n) = 2(B_{\phi}(0,0)-B_{\phi}(m,n))
\end{equation}
where $B_{\phi}(m,n)$ is the phase autocorrelation matrix and $(m,n)$ are the coordinates along x and y-axis.
The 2D phase autocorrelation matrices for the FFT-based phase screen and the modified subharmonic method by Johansson and Gavel \cite{Johansson94} are represented as follows.

\begin{equation} \label{eq:eq2}
B_{\phi}^{FFT}(m, n) = \sum_{m^{\prime}=-N_{x} / 2}^{N_{x} / 2-1} \sum_{n^{\prime}=-N_{y} / 2}^{N_{y} / 2-1} f^{2}_{FFT}\left(m^{\prime}, n^{\prime}\right) e^{i 2 \pi\left(\frac{m^{\prime} m}{N_{x}}+\frac{n^{\prime} n}{N_{y}}\right)} 
\end{equation}

\begin{equation} \label{eq:eq3}
B_{\phi}^{SUB}(m, n)=\sum_{p=1}^{N_{p}} \sum_{m^{\prime}=-3}^{2} \sum_{n^{\prime}=-3}^{2} f_{SUB}^{2}\left(m^{\prime}, n^{\prime}\right) e^{i 2 \pi 3^{-p}\left(\frac{\left(m^{\prime}+0.5\right) m}{N_{x}}+\frac{\left(n^{\prime}+0.5\right) n}{N_{y}}\right)} 
\end{equation}
where $f^{2}_{FFT}\left(m^{\prime}, n^{\prime}\right)$ and $f_{SUB}^{2}\left(m^{\prime}, n^{\prime}\right)$ are the  \vk\ spectrum and subharmonic power spectrum as explained by Johansson and Gavel. $(N_x,N_y)$ are sample points, p is the $p^{th}$ subharmonic and $N_p$ is the total number of subharmonics. Set $f_{FFT}$= 0, for $(m^{'},n^{'})= (0,0)$ and $f_{SUB}$ = 0, for $(m^{'},n^{'})= (-1,0)$ and $(0,-1)$ as originally proposed by Lane et. al\cite{Lane92}. There will be an overlap between subharmonic energy sample and secondary lobes from first sample of high frequency spectrum or harmonic sample during subharmonic addition. Earlier this leakage of energy has been dealt using patch normalization factor, where first patch of high frequency spectrum is weighted by 0.707  for $(m^{'},n^{'})= (\pm 1,0)$ and $(m^{'},n^{'})= (0,\pm 1)$ and 0.866 for $(m^{'},n^{'})= (\pm 1,\pm 1)$ in the original method of Johansson and Gavel\cite{Johansson94}. Similarly, the original method of Lane et al.\cite{Lane92}, Sedmak\cite{Sedmak04} proposed the corresponding weights to be 0.935 and 0.998 respectively. Our simulations show that these weights do not fit perfectly for different $G/L_0$ ratios and hence need to be tuned on a case by case basis. We have made our approach independent from these weights assignments. The weight factor has been set equal to 1 in our approach. Section ~\ref{sec:algorithm} explains this approach in detail.\\ 
The 2D phase autocorrelation matrix after compensating with subharmonics is represented as
    \begin{equation}\label{eq:eq4}
    B_{\phi}(m,n) = B^{FFT}_{\phi}(m,n) +B^{SUB}_{\phi}(m,n)
    \end{equation}
    
\section{Algorithm to compensate for residual error in phase structure function}
\label{sec:algorithm}
To calculate the remaining error in the final $B_{\phi}(m,n)$, eq.~(\ref{eq:eq4}) is converted to phase structure matrix $D_{\phi}(m,n)$ with the help of eq.~(\ref{eq:eq1}) with the assumption that $B_{\phi}^{FFT}(0,0)$ and $B_{\phi}^{SUB}(0,0)$ are zero because we are not concerned about the piston component. This gives the following equation
\begin{equation}\label{eq:eq5}
        D_{error}(m,n) = D_{theory}(m,n) - D_{\phi}(m,n)  
    \end{equation}
where $D_{theory}(m,n)$ is the well-known theoretical \vk\ phase structure matrix \cite{Johansson94}, given as follows:

\begin{equation}\label{eq:eq6}
    D_{theory}(r)= 6.16 r_0^{5/3}\Bigg[0.6(L_0/2\pi)^{5/3} - \frac{(rL_0/4\pi)^{5/6}}{\gamma(11/6)}K_{5/6}(2\pi
    r/L_0)\Bigg]
\end{equation}
where $r^2=(m\Delta)^2 + (n\Delta)^2$, $\Delta=G/N$.

We need to compensate $D_{\phi}$ so that $D_{error}$ is minimized. However, simply adding error correction terms in the $D_{\phi}$ matrix directly would only introduce further error into the system, while taking the Fourier transform. This is because any matrix or curve in general will have higher order moments. Thus, if we take the Fourier transform of the adjusted equation, the resultant curve will have completely different moments and hence power spectrum. This is because the transition between two steps in the error matrix will not be smooth, which introduces additional errors due to Gibb's phenomena like overshoots. Just curve fitting with any function does not satisfy the additional requirement of leaving the power spectrum unaffected by the process. What we really need is to introduce a smoothening operator like a Gaussian function in the phase autocorrelation matrix which exactly compensates for $D_{error}$.

For that we have developed an iterative algorithm (see the flow chart shown in Fig.~\ref{fig:flow}) and implemented it in Matlab. The algorithm looks for the perfect Gaussian curve that minimizes the $D_{error}$ matrix. We use Matlab cftool to initially determine the correct 1D Gaussian matrix and later convert it into a 2D matrix by exploiting the fact that $B_\phi$(r), $B_{theory}$(r) and $B_{SUB}$(r) all are dependent on r only and hence are centre symmetric functions. We call the fitted Gaussian phase structure matrix $D_{gauss}$ and the corresponding Gaussian phase autocorrelaiton matrix $B_{gauss}$ (using eq.~(\ref{eq:eq1})).

\begin{figure}[h!]
    \centerline{\includegraphics[width=0.8\textwidth]{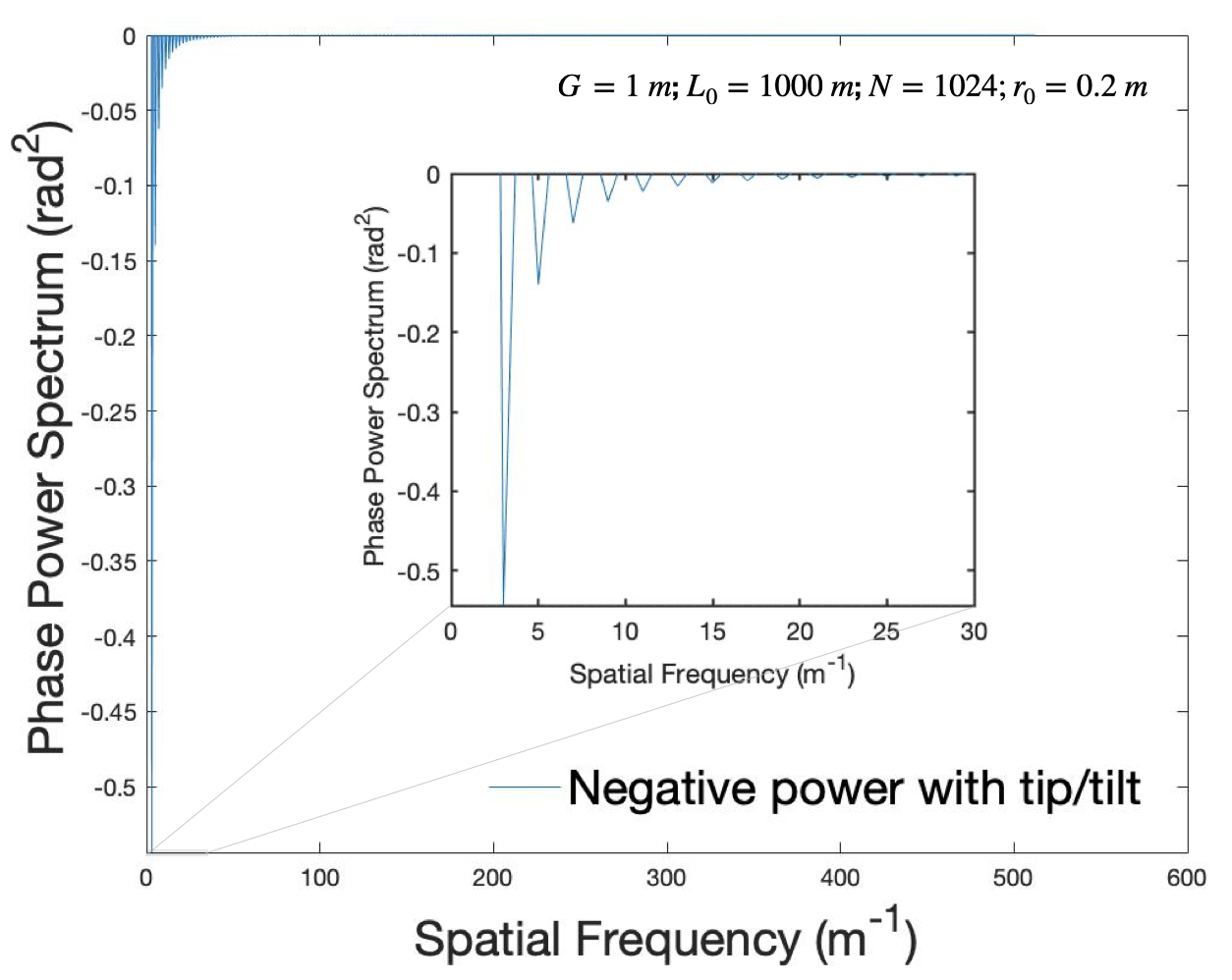}}
    \caption{\label{fig:power_withtiptilt}Negative power spectrum values for small $G/L_0$ ratios}
\end{figure}

The final equation for $B_{tot}$ can then be written as 
\begin{equation}\label{eq:eq7}
    B_{tot}(m,n) = B^{FFT}_{\phi}(m,n) +B^{SUB}_{\phi}(m,n) - D_{gauss}(m,n)/2
    \end{equation}
Here we have used $B_{gauss} = - D_{gauss}/2$ from eq.~(\ref{eq:eq1}). A look at  the power spectrum of $B_{tot}(m,n)$ in Fig~\ref{fig:power_withtiptilt} shows that it  contains negative terms \cite{Xiang14} for the case of $G/L_0<1$. Directly putting those frequency terms equal to zero leads to a loss in the energy spectrum. Hence $B_{tot}$(m,n) matrix needs to be preprocessed to eliminate most of these negative values in the power spectrum. Over small frequencies, piston and tip/tilt components account for most of these high magnitude negative elements. Therefore, we first extract the piston and tip/tilt components from the phase autocorrelation matrix $B_{tot}$. The tip/tilt component from phase autocorrelation matrix is given as\cite{Xiang14}

\begin{figure}[h!]
    \centerline{\includegraphics[width=0.8\textwidth]{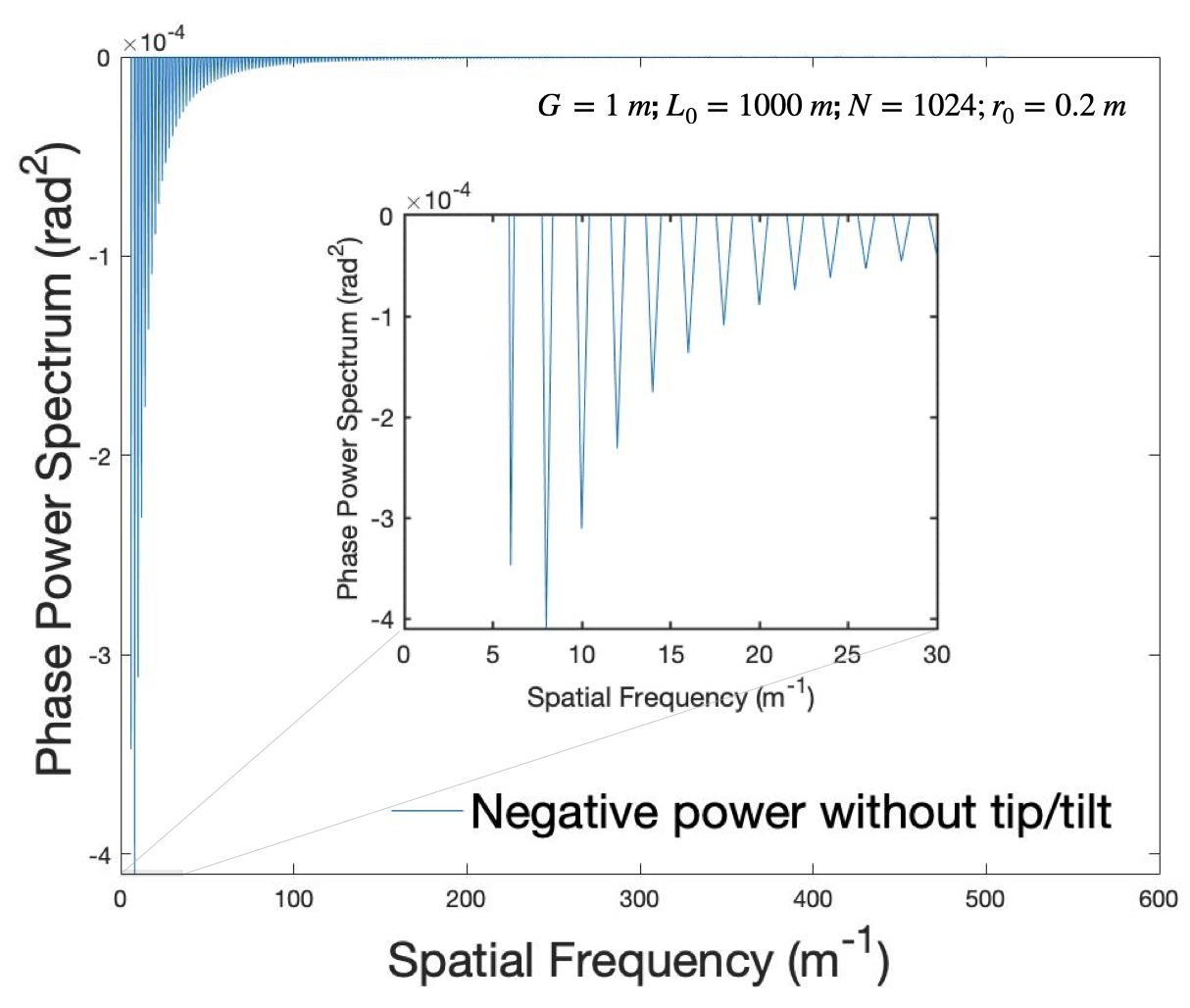}}
    \caption{\label{fig:power_withouttiptilt}Residual negative power spectrum values after removing tip/tilt from $B_{tot}$ for small $G/L_0$ ratios}
\end{figure}
\begin{equation}\label{eq:eq8}
            B_{tilt}(r) = B_{tilt}(0) - r^2\sigma_{tilt}^2/2 
        \end{equation}
where $\sigma^2_{tilt}$ is the variance of the random tilt angle in the x or y directions and given as follows\cite{Xiang14}: 

\begin{equation}\label{eq:eq9}
            \sigma^2_{tilt} = \frac{B_{tot}(G/2+\Delta)-B_{tot}(G/2)}{\Delta(G-\Delta)/2}   
\end{equation}
After setting, $B_{tilt}(0)$ = 0, the remaining phase autocorrelation matrix is given as follows:
\begin{equation}\label{eq:eq10}
            B_{high}(r) = B_{tot}(r) - B_{tilt}(r)
\end{equation}
The power spectra $f_{high}^2$ and $f_{tilt}^2$ of the phase autocorrelation matrices $B_{high}(r)$ and $B_{tilt}(r)$ are obtained by standard fourier transformation. 
Fig.~\ref{fig:power_withouttiptilt}, shows the remaining negative power elements present in the power spectrum of $B_{high}$ matrix. In comparison to Fig.~\ref{fig:power_withtiptilt}, the largest negative power contributions fall by factor of three order of magnitude. Now we set the negative values in $f_{high}^2$ equal to zero by hand. The new error matrix is given as:
\begin{equation}\label{eq:eq11}
            B_{high}^{err}(r) = B_{high}^{'}(r) - B_{high}(r)  
\end{equation}
where $B_{high}^{'}(r)$ is the phase autocorrelation matrix obtained after setting the negative elements in $f_{high}^2$ to zero. The residual error that is present in the high frequency region can then be reduced with the help of a Gaussian smoothing operator, using Matlab fmincon tool. The high frequency compensated matrix is given as :
\begin{equation}\label{eq:eq12}
            B_{high}^{comp}(r) = B_{high}(r) - H_{high}^{comp}(r)B_{high}^{err}(r)  
\end{equation}
where $H_{high}^{comp}(r)$ is the smoothening operator, multiplied with error matrix to reduce the high frequency errors. fmincon gives the optimised parameter for smoothening operator by calculating the final error in the $D_{\phi}(r)$ matrix w.r.t. $D_{theory}(r)$.

\section{Implementation of the compensation algorithm}\label{sec:flow_chart}
In this section, we explain the error compensation algorithm with the help of the flow chart shown in Fig.~\ref{fig:flow}. Brief explanations of each of the steps from \emph{$L_1$} to \emph{$L_{12}$} are given below. 
\begin{figure}[h!]
    \centerline{\includegraphics[width=1.0\textwidth]{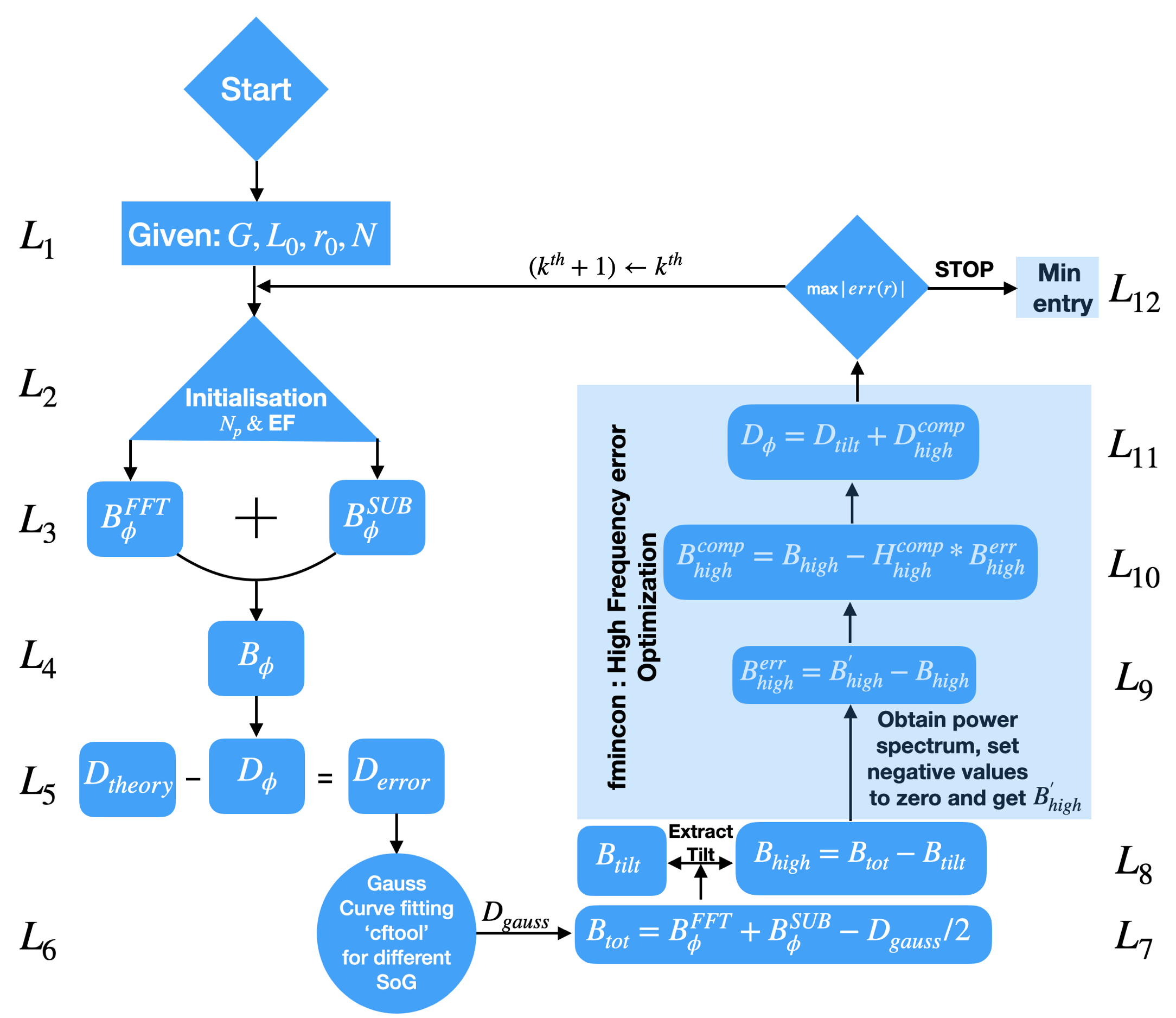}}
    \caption{\label{fig:flow}Flow chart for error compensation}
\end{figure}

\begin{itemize}

\item[$L_1$:] Input screen size $G$, outer scale size $L_0$, Fried parameter $r_0$, and number of samples $N$.

\item[$L_2$:] Initialize the algorithm with $N_p$, the total number of subharmonics and extrapolation factor EF, both ranging from 1 to 10. The EF factor is relevant while performing curve fitting. e.g. EF = 2 means curve fitting will work from 3 to N/2 points and later that curve will be extrapolated from 1 to N/2 points ( example shown in section \ref{sec:example}, Fig.~\ref{fig:1D_error} )

\item[$L_3$:] Obtain $B_{\phi}^{FFT}$ and $B_{\phi}^{SUB}$ based upon \emph{$L_1$} and \emph{$L_2$} parameters.

\item[$L_4$:] Add the matrices that were calculated in \emph{$L_3$} layer, call that $B_{\phi}$.

\item[$L_5$:] Obtain $D_{\phi}$ from $B_{\phi}$ using eq.~(\ref{eq:eq1}) and also produce $D_{theory}$ matrix based upon \emph{$L_1$} and \emph{$L_2$}. Then obtain error matrix $D_{error}$ using eq.~(\ref{eq:eq5}).

\item[$L_6$:]  Obtain 1D array from $D_{error}$ matrix from the centre and give as input to the curve fitting tool cftool  which works on 1D data. The output of the tool will be a best fitted curve in terms of Sum of Gaussian's (SoG), called $D_{gauss}$

\item[$L_7$:] The final expression for $B_{tot}$ is shown eq.~(\ref{eq:eq7}).

\item[$L_8$:] Extract tilt component from $B_{tot}$ matrix using eq.~(\ref{eq:eq8}) and eq.~(\ref{eq:eq9}).

\textbf{fmincon}: High Frequency error Optimization

\item[$L_9-L_{11}$:] Obtain error matrix $B_{high}^{err}$ after setting negative elements in the power spectrum to zero. Compensate for high frequency error by multiplying error matrix with smoothening operator-SoG. Calculate the maximum remaining error in structure function matrix relative to the $D_{theory}$ matrix. Thus fmincon will give parameters for the smoothing operator, that gives the lowest possible residual error.

\item[$L_{12}$:] Update the entry for $N_p$ $\&$ EF to next value, evaluation from $L_2$-$L_{12}$ would go in loop, and MRE value stored in vector form. At the end of the iterations, min entry will get extracted out from stored vector and accepted for final analysis. 
\end{itemize}

Table~\ref{tab:t1} shows the result of curve fitting using cftool for different cases of $G/L_0$ and N, which demonstrates that the Gaussian error matrix can compensate for a wide range of $G/L_0$ ratios and under different sampling constraints.

\begin{table}[]
\resizebox{\textwidth}{!}{%
\begin{tabular}{|c|c|c|c|c|c|c|c|c|c|c|c|c|c|c|c|c|c|}
\hline
\multicolumn{2}{|c|}{} & \multicolumn{4}{c|}{$N$ = 128} & \multicolumn{4}{c|}{$N$ = 256} & \multicolumn{4}{c|}{$N$ = 512} & \multicolumn{4}{c|}{$N$ = 1024} \\ \hline
$G$       & $L_0$      & $N_p$  & $EF$  & SoG & MRE(\%) & $N_p$  & $EF$  & SoG & MRE(\%) & $N_p$  & $EF$  & SoG & MRE(\%) & $N_p$  & $EF$  & SoG  & MRE(\%) \\ \hline
1         & 20         & 2      & 1     & 3   & 0.25    & 2      & 2     & 4   & 0.61    & 2      & 2     & 3   & 0.36    & 2      & 1     & 4    & 1       \\ \hline
5         & 20         & 1      & 1     & 5   & 0.19    & 6      & 3     & 5   & 0.23    & 4      & 2     & 3   & 0.27    & 7      & 1     & 3    & 1       \\ \hline
10        & 20         & 9      & 5     & 4   & 0.25    & 9      & 1     & 5   & 0.29    & 1      & 1     & 3   & 0.26    & 7      & 3     & 6    & 0.35    \\ \hline
20        & 20         & 3      & 1     & 4   & 0.53    & 10     & 1     & 6   & 0.4     & 4      & 5     & 5   & 0.97    & 3      & 5     & 3    & 0.22    \\ \hline
40        & 20         & 5      & 0     & 4   & 0.95    & 8      & 0     & 6   & 0.75    & 8      & 0     & 3   & 1       & 7      & 1     & 3    & 2.99    \\ \hline
60        & 20         & 5      & 0     & 4   & 0.17    & 3      & 2     & 6   & 0.42    & 9      & 8     & 5   & 0.41    & 3      & 1     & 5    & 0.94    \\ \hline
80        & 20         & 8      & 1     & 4   & 0.24    & 7      & 5     & 3   & 0.17    & 5      & 6     & 3   & 0.32    & 6      & 4     & 3    & 0.3     \\ \hline
100       & 20         & 8      & 1     & 4   & 0.24    & 7      & 5     & 3   & 0.17    & 5      & 6     & 3   & 0.32    & 6      & 4     & 3    & 0.3     \\ \hline
1         & 10         & 3      & 1     & 4   & 0.53    & 10     & 1     & 6   & 0.4     & 4      & 5     & 5   & 0.97    & 3      & 5     & 3    & 0.22    \\ \hline
1         & 100        & 3      & 1     & 6   & 0.21    & 4      & 1     & 4   & 0.49    & 3      & 6     & 5   & 0.96    & 3      & 6     & 5    & 0.25    \\ \hline
1         & 1000       & 5      & 0     & 4   & 0.95    & 8      & 0     & 6   & 0.75    & 8      & 0     & 3   & 1       & 7      & 1     & 3    & 2.99    \\ \hline
10        & 100        & 2      & 0     & 6   & 0.28    & 3      & 0     & 4   & 0.22    & 8      & 1     & 4   & 0.30    & 7      & 0     & 3    & 0.27    \\ \hline
10        & 1000       & 5      & 0     & 4   & 0.17    & 3      & 2     & 6   & 0.42    & 9      & 8     & 5   & 0.41    & 3      & 1     & 5    & 0.94    \\ \hline
100       & 100        & 3      & 0     & 6   & 0.21    & 2      & 3     & 5   & 0.21    & 7      & 0     & 3   & 0.26    & 1      & 1     & 3    & 0.26    \\ \hline
100       & 1000       & 8      & 1     & 4   & 0.24    & 7      & 5     & 3   & 0.17    & 5      & 6     & 3   & 0.32    & 6      & 4     & 3    & 0.3     \\ \hline
\end{tabular}%
}
\caption{Result of curve fitting against Gaussian function for different cases of $G/L_0$ and N in terms of Maximum Relative Error (MRE) for fixed $r_0$ = 0.2 m}
\label{tab:t1}
\end{table}

\subsection{Example}\label{sec:example}
To illustrate the robustness of the above algorithm, we have taken an example with G = 80 m, say for a large future telescope, N = 256, and median value of $L_0$ = 20 m. 

\begin{figure}[h!]
    \centerline{\includegraphics[width=0.8\textwidth]{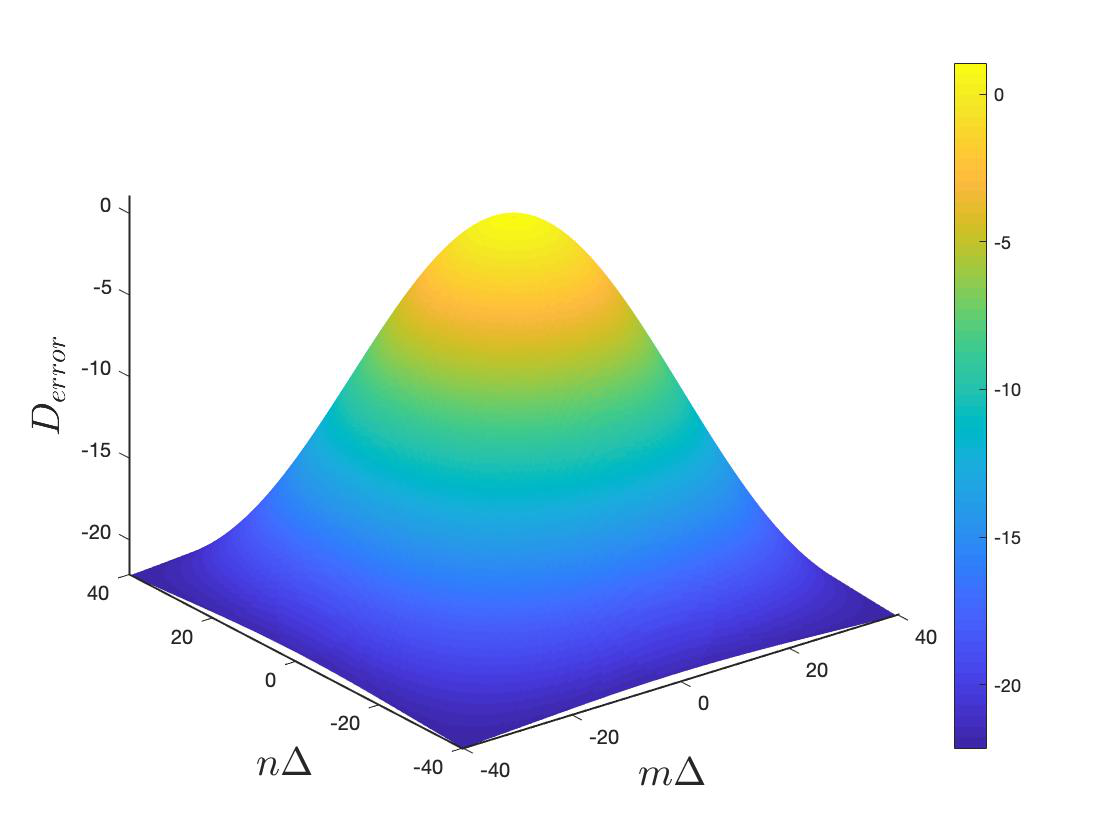}}
    \caption{\label{fig:3d_error} 3D $D_{error}$ matrix for case $G$ = 80 m, $L_0$ = 20 m , $r_0$ = 0.2 m, $N_x$=$N_y$= 256 }
\end{figure}

The output from the above algorithm corresponding to minimum error entry as in (step $L_{12}$), has been plotted against $EF$ = 5 and $N_p$ = 8. Fig.~\ref{fig:3d_error} gives a 3D rendering of $D_{error}$ matrix with a maximum separation of up to 40 m, corresponding to eq. (\ref{eq:eq4}). 

\begin{figure}[h!]
    \centerline{\includegraphics[width=0.9\textwidth]{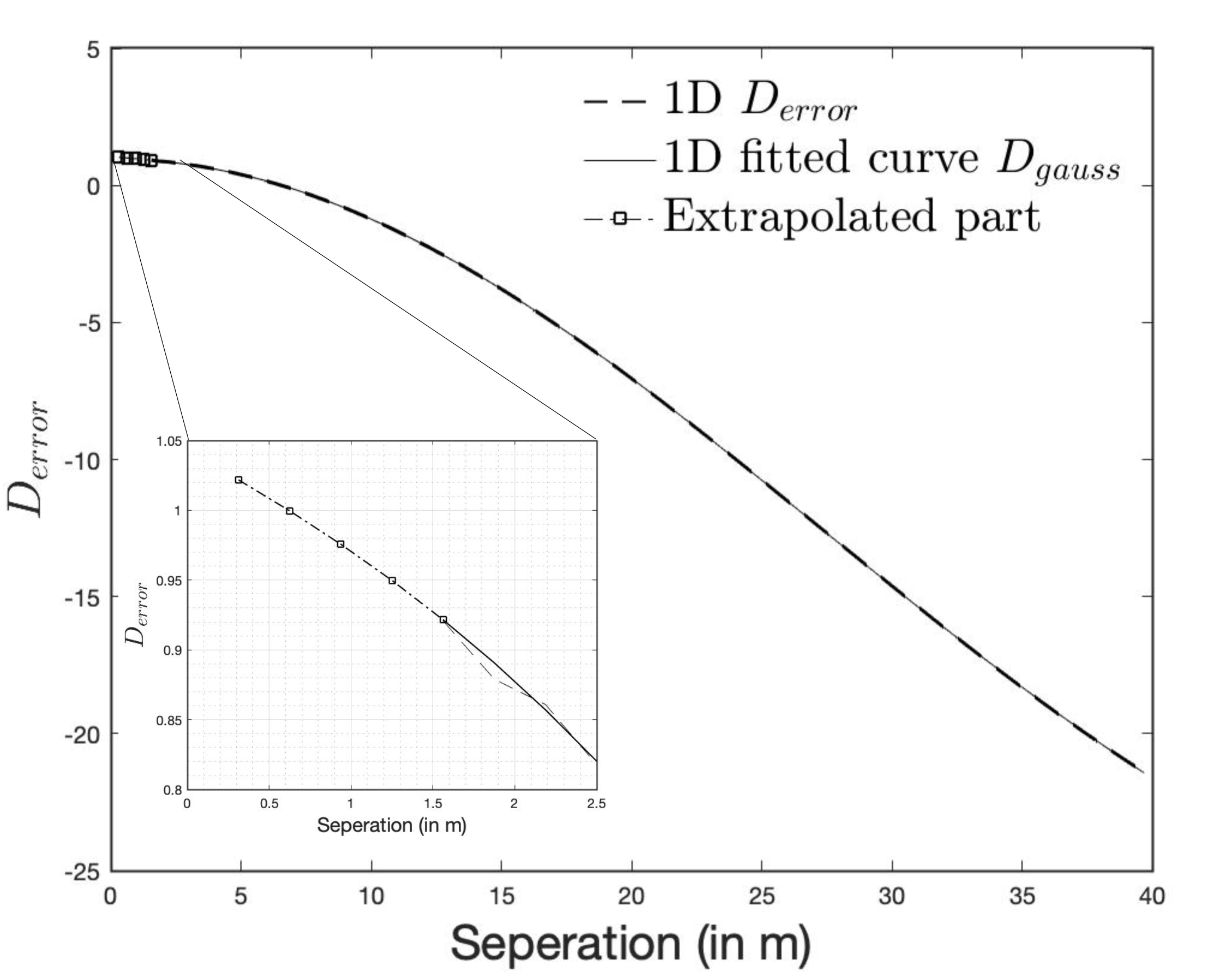}}
    \caption{\label{fig:1D_error} 1D $D_{error}$ matrix fitted against $D_{gauss}$ matrix , along with an extrapolated part of the curve. Here $G$ = 80 m, $L_0$ = 20 m , $r_0$ = 0.2 m, $N_x$=$N_y$= 256}
\end{figure}

Fig.~\ref{fig:1D_error} represents 1D $D_{error}$ matrix ( radial section from 3D $D_{error}$ matrix ) along with 1D fitted curve $D_{gauss}$ including the extrapolated part, for a maximum separation of up to 40 m.

\begin{figure}[h!]
    \centerline{\includegraphics[width=0.8\textwidth]{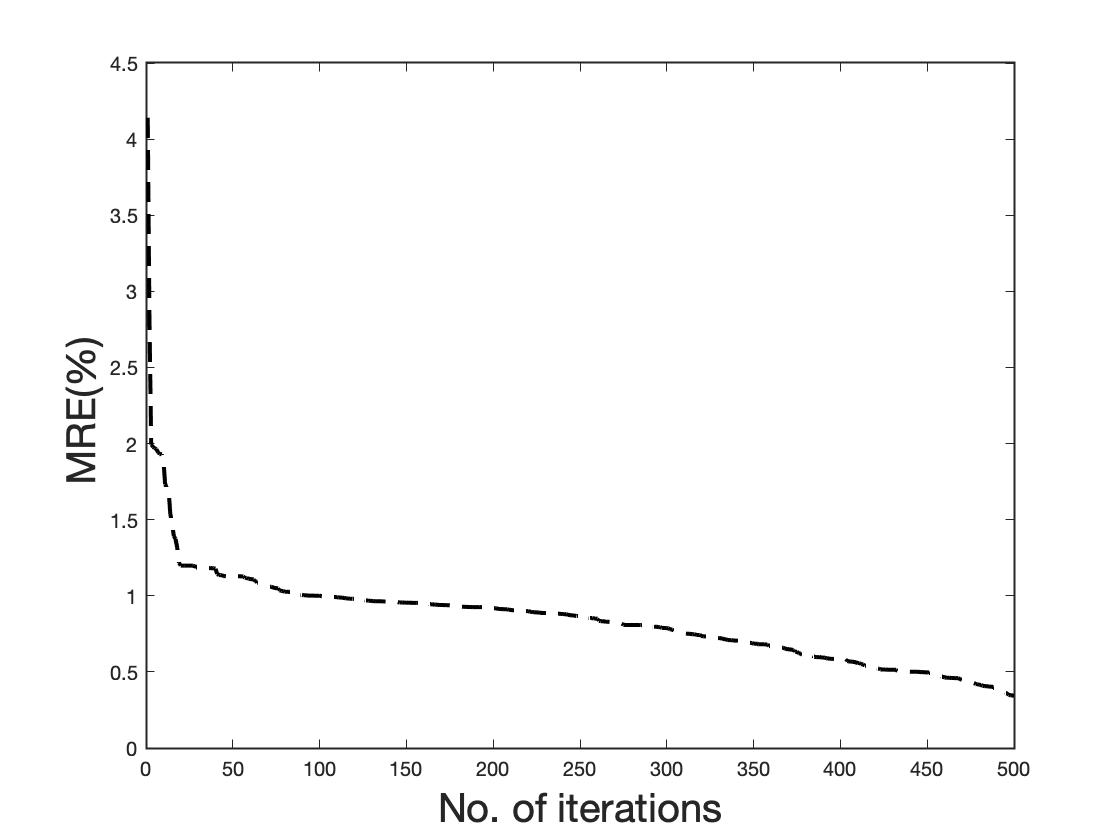}}
    \caption{\label{fig:mre_iterations}Maximum relative error MRE with the maximum number of iterations for $G$ = 80 m, $L_0$ = 20 m , $r_0$ = 0.2 m, $N_x$=$N_y$= 256 }
\end{figure}

Lastly, MRE values are stored against 500 entries corresponding to $N_p$ ranging from 1 to 10, $EF$ ranging from 1 to 10 and $SoG$ ranging from 2 to 6 after performing cftool fitting. This has been arranged in descending order and presented in Fig.~\ref{fig:mre_iterations}, which illustrates a large set of iterations where errors are less than 1\% and entry with minimum MRE has been picked up. Typical time required to perform each iteration for this case is $\approx$4.9 sec on 2.3GHz quad-core Intel Core i5 Macbook pro 2018 model. 

\section{Validation \lowercase{via} phase structure function calculated from simulated phase screen}
\label{sec:Validation} 
 To obtain the phase screen $\phi(m,n)$ from the power spectrum, the following relation is used\cite{Xiang14}:
\begin{equation}\label{eq:eq13}
            \begin{aligned} 
            \phi(m \Delta, n \Delta)=& \sum_{m^{\prime}=-N / 2}^{N / 2-1} \sum_{n^{\prime}=-N / 2}^{N / 2-1}\left[R_{a}\left(m^{\prime}, n^{\prime}\right)+\mathrm{i} R_{b}\left(m^{\prime}, n^{\prime}\right)\right] f\left(m^{\prime} \Delta^{\prime}, n^{\prime} \Delta^{\prime}\right) \exp \left[\mathrm{i} 2 \pi\left(m^{\prime} m+n^{\prime} n\right) / N\right] 
            \end{aligned}
\end{equation}
where $R_a(m^{'},n^{'})$ and $R_b(m^{'},n^{'})$ are zero-mean and unity-variance gaussian random number generator. We get $\phi_{high}$ and $\phi_{tilt}$, by replacing $f$ with $f_{high}$ and $f_{tilt}$, which are square roots of the power spectrum corresponding to autocorrelation matrix $B_{high}^{comp}$ and $B_{tilt}$  respectively.

\begin{figure}[h!]
    \centerline{\includegraphics[width=0.9\textwidth]{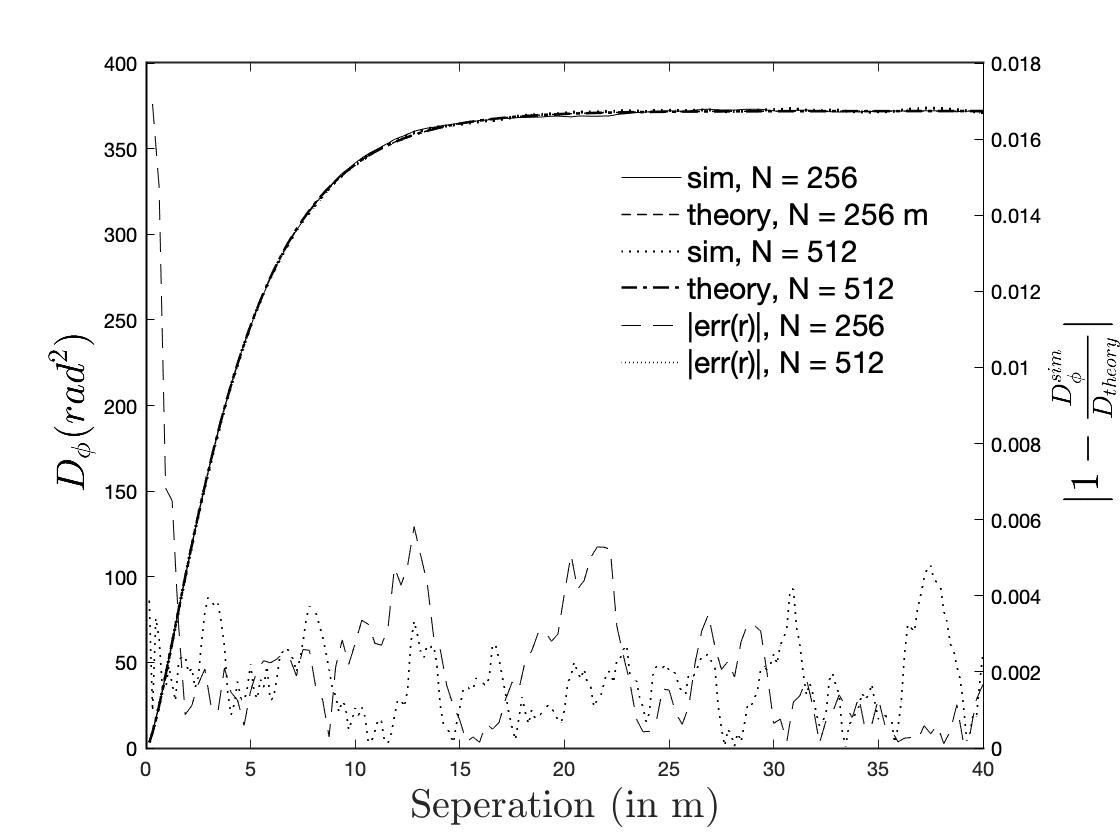}}
    \caption{\label{fig:large} Left: Compares simulated structure function w.r.t theoretical structure function for maximum separation of $G/2$ for two different cases $N$ = 256 and 512, for fixed $G$ = 80 m, $r_0$ = 0.2 m and $L_0$ = 20 m. Right:  Calculates the magnitude of relative error in simulated structure function for maximum separation of $G/2$, for both the cases.}
\end{figure}

For validation, we consider scenarios of apertures up to 40 m  i.e $G$ = 80 m, at a median $L_0$ = 20 m for two different sampling levels N = 256 and 512. The phase structure function, defined as an ensemble average of differences of phases at various separation\cite{roddier1999adaptive}, has been averaged over 100K independent frames. The relative error in phase structure function is calculated as follows:

\begin{equation}\label{eq:eq14}
       err(r)= \frac{D_{\phi}^{sim}(r)-D_{theory}(r)}{D_{theory}(r)}
\end{equation}
Here, $D_{\phi}^{sim}(r)$ is the phase structure function from the simulated phase screen. The magnitude of the peak relative error $max(\vert err(r) \vert)$ is $<1.6\%$ for $N$ = 256 and $<0.5\%$ for $N$ = 512 as shown in Fig. \ref{fig:large}.

\begin{figure}[h!]
    \centerline{\includegraphics[width=0.8\textwidth]{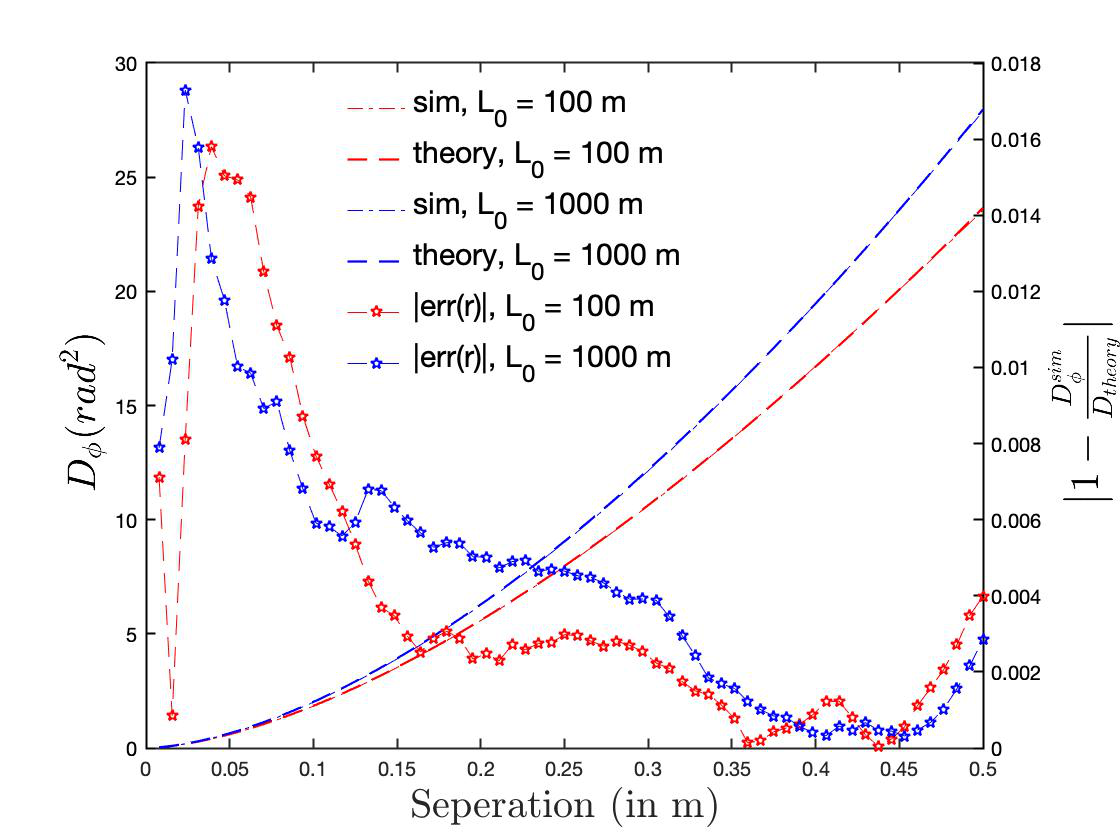}}
    \caption{\label{fig:large_L0} Left: Compares simulated structure function w.r.t theoretical structure function for maximum separation of $G$/2 for two different cases $L_0$ = 100 m and 1000 m , for fixed $G$ = 1 m, $r_0$ = 0.2 m and $N$ = 128 . Right: Calculates the magnitude of relative error in simulated structure function for maximum separation of $G$/2, for both the cases.}
\end{figure}

\begin{figure}[h!]
    \centerline{\includegraphics[width=0.99\textwidth]{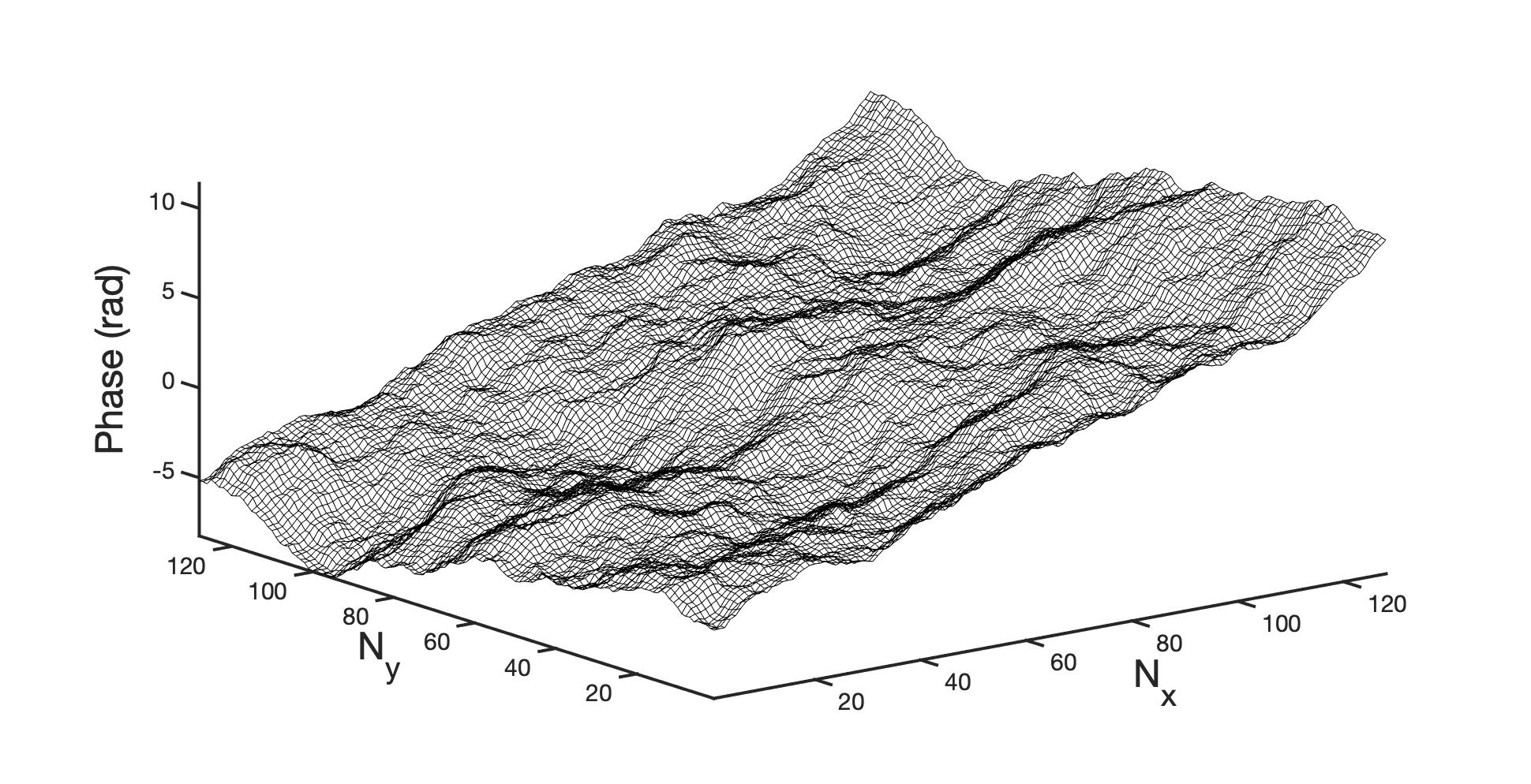}}
    \caption{\label{fig:screen}Phase Screen for case $G$ = 1 m, $L_0$ = 100 m , $r_0$ = 0.2 m, $N_x$=$N_y$= 128 }
\end{figure}

We also illustrate the performance ( shown in Fig. \ref{fig:large_L0} ) with parameters G = 1 m, $L_0$ = 100 m and 1000 m, N = 128, $r_0$ = 0.2 m  which cover the extreme cases ( very low $G/L_0$ ratios ) which leads to the maximum error in the simulation. The magnitude of the peak relative error $max(|err(r)|)$ is $< 1.6\%$ for $L_0$ = 100
m and $< 1.8\%$ for $L_0$ = 1000 m.  Fig.~\ref{fig:screen} shows one realization of the corresponding phase screen plots for $L_0$ = 100 m.

\begin{figure}[h!]
    \centerline{\includegraphics[width=0.75\textwidth]{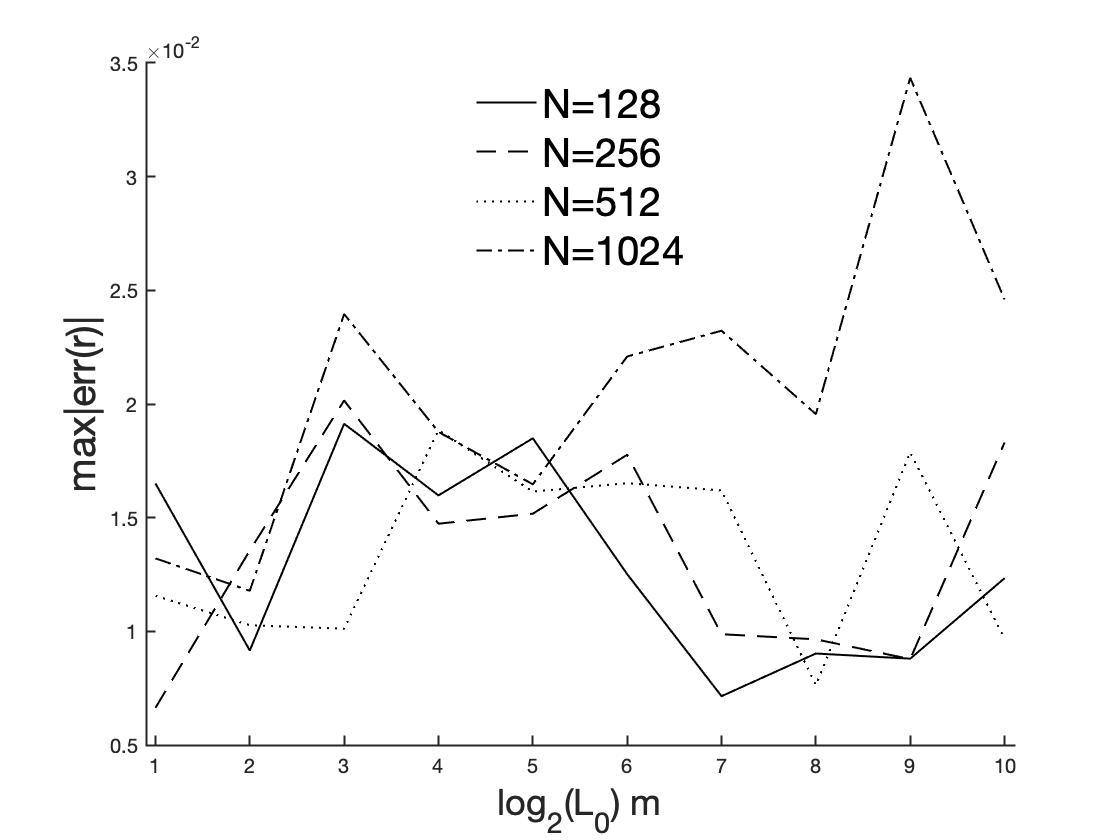}}
    \caption{\label{fig:error_L}The magnitude of the peak relative error for N = 128, 256, 512 and 1024 for $L_0$ ranges up to 1024 m. Here G = 2 m, $r_0$ = 0.2 m.  }
\end{figure}

\begin{figure}[h!]
    \centerline{\includegraphics[width=0.75\textwidth]{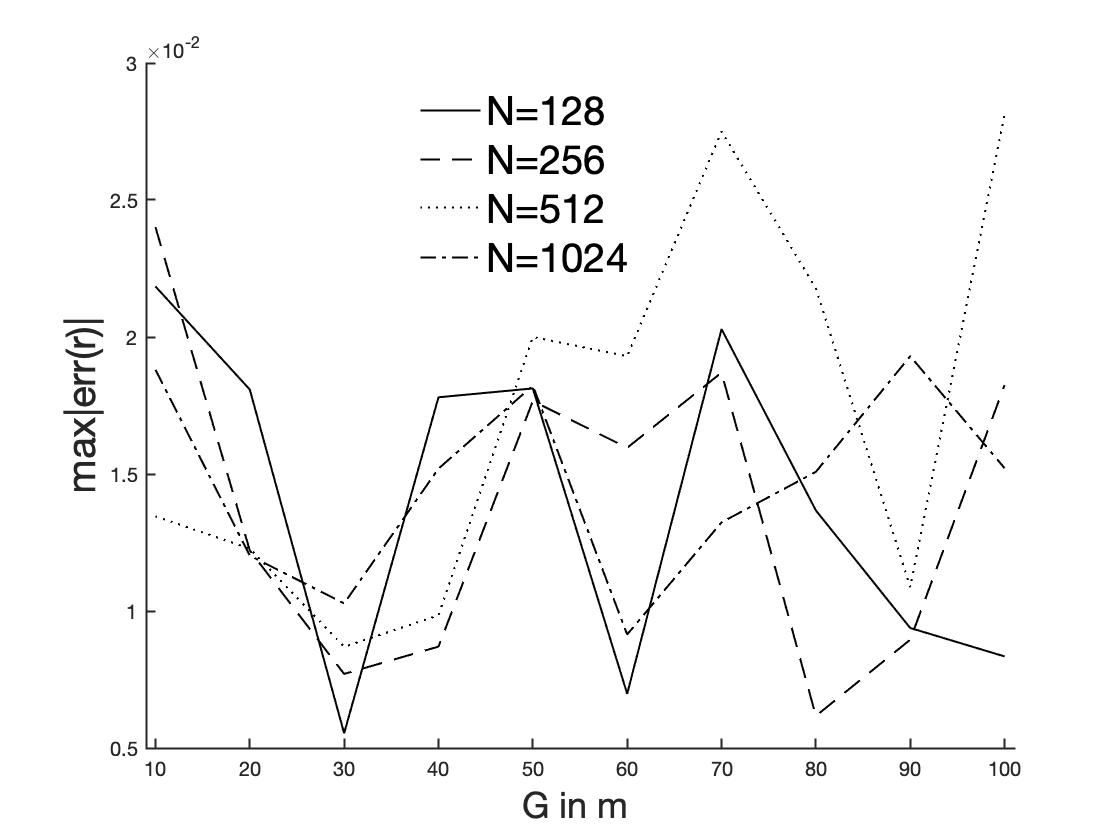}}
    \caption{\label{fig:error_G}The magnitude of the peak relative error for N = 128, 256, 512 and 1024 for screen size up to 100 m. Here $L_0$ = 25 m, $r_0$ = 0.2 m.}
\end{figure}

Fig.~\ref{fig:error_L} contains results of magnitude of the peak relative error in $D_{\phi}^{sim}(r)$ for the case of different sampling points $N$ = 128/256/512/1024, for $L_0$ ranges up to 1024 m, $r_0$ = 0.2 m and $G$ = 2 m. Similarly, Fig.~\ref{fig:error_G} contains results of magnitude of the peak relative error in $D_{\phi}^{sim}(r)$ for the case of different sampling points $N$ = 128/256/512/1024, for $G$ ranges up to 100 m, $r_0$ = 0.2 m and $L_0$ = 25 m.


There are some outliers which have a high residual error as shown in Fig.~\ref{fig:error_L} and Fig.~\ref{fig:error_G}, because we have not set the phase autocorrelation matrix to zero for $r>G/2$. The reason for this stems from the non-zero value of $B_{high}(r)$ at $r>G/2$, where $B_{high}(r)$ is formed from the removal of piston and tilt from $B_{tot}(r)$. This can be resolved by using a better smoothening operator, which we can multiply with $B_{high}$ so that it falls to zero progressively and not sharply. This can provide further improvement in the compensation.

\section{Conclusion}
\label{sec:Conclusion} 
In this paper, we put forward a new method to compensate for the residual error in both the Low and/or High-frequency region of FFT simulated phase screens that remain even after compensating with the modified subharmonic method. This method provides accurate phase screen structure for even $G/L_0$ ratios as small as 1/1000 plus screen sizes as large as 100 m. No Patch Normalization factor is needed, no need to calculate subharmonic weight coefficient \cite{Lane92} and weights to compensate for high-frequency components, as done by Sedmak\cite{Sedmak04}.  While adequately large $G/L_0$ ratios may be the natural choice for modern large telescopes, simulations that deal with applications such as laser beam propagation through turbulent atmospheres would tend to have very small $G/L_0$ ratios. The method we propose is independent of the $G/L_0$ ratio choice. However, we emphasize that properly sampling $r_0$ and the high-frequency phase spectrum forces $N$ to be at least larger than $(2G/r_0)$ and preferably up to the inner scale limit $(2G/l_0)$. Currently we have demonstrated this technique for only circular screens. We have used a GPU processor with total number of 128 cores, such that each iteration runs independently on each core. We have fixed the number of iterations to 500, although increasing this will lead to improvement of errors in some cases. Each core takes about $\sim$0.06, $\sim$0.1, $\sim$0.36 and $\sim$1.1 minutes for sampling sizes of 128, 256, 512 and 1024. On the above GPU system, this translates to total computing times for error minimization of about $\sim$0.2, $\sim$0.5, $\sim$1.25 and $\sim$4 minutes for sampling sizes of 128, 256, 512 and 1024 respectively. Once the coefficients are determined, generating multiple phase screen realizations from the corresponding power spectrum takes a few milliseconds at most on 2.3 GHz quad-core Intel Core i5 Macbook pro 2018 model. Then it takes less than a minute to $\sim$10 min for averaging over 100k phase screens, for sampling sizes ranging from 128 to 1024.

The uniqueness of our approach is its ability to deal with any $G/L_0$ ratio within a very broad range, in an automated iterative process with little human intervention needed for tuning of parameters. Any standard FFT based approach (say Sedmak's\cite{Sedmak04} compensated approach) for a given computer platform is computationally fast, only if we already have determined proper measures of the various compensating components such as the patch normalization factor, subharmonic weights, high-frequency weights etc. Typically, determining these compensations is where the difficulty is due to mathematical complexity, algorithmic limitations and/or computational power requirements. Our algorithm accomplishes the determination of the required compensation in very little time, with fairly reasonable computational power while at the same time keeping the residual errors competitively low by using an appropriate compensator. Other existing FFT based approaches have limitations in their operable $G/L_0$ range. For example, Xiang et. al.\cite{Xiang14} offer a very computationally fast approach but does not apply subharmonic compensation. Zhang et. al.\cite{Zhang19} does not consider compensation for the high-frequency error, thus leaving a residual error of more than 100\% in the high-frequency region. Sedmak's\cite{Sedmak04} approach needs the determination of accurate subharmonic weights for different $G/L_0$ ratios. The accuracy of our method from low-frequency to the high-frequency range is between 0.5-3\% for $G/L_0$ as low as 1/1000 and screen size up to 100 m in diameter.

\acknowledgments 
 We would like to thank Sedmak for providing insights into the nature of atmospheric phase power spectrum through private communication. We also thank Xiang for sharing his MATLAB code which calculates the phase structure function quickly for a large number of phase screens. We acknowledge usage of IUCAA's Pegasus cluster computer for running multiple independent iterations in parallel.

\bibliography{report} 
\bibliographystyle{spiejour}   


\vspace{2ex}\noindent\textbf{Sorabh Chhabra} is a PhD student at Inter University Center for Astronomy and Astrophysics (IUCAA), Pune. He received his B.Tech degree in Electronics and Communication from Delhi Technological University (formally Delhi College of Engineering ) in 2016 and joined for his PhD in the same year in Instrumentation department at IUCAA.

\vspace{1ex}

\listoffigures
\listoftables

\end{spacing}

\end{document}